**Title:**

# Determination of the He-He, Ne-Ne, Ar-Ar, and H$_2$ interaction potential by wave function imaging

**Authors:** S. Zeller[1]*, M. Kunitski[1], J. Voigtsberger[1], M. Waitz[1], F. Trinter[1], S. Eckart[1], A. Kalinin[1], A. Czasch[1], L. Ph. H. Schmidt[1], T. Weber[2], M. Schöffler[1], T. Jahnke[1], and R. Dörner[1]*

**Affiliations:**

[1]Institut für Kernphysik, Goethe-Universität Frankfurt am Main, Max-von-Laue-Strasse 1, 60438 Frankfurt am Main, Germany

[2]Chemical Sciences Division, Lawrence Berkeley National Laboratory, Berkeley, CA 94720, USA

*Corresponding author.

## Abstract

We report on a direct method to measure the interatomic potential energy curve of diatomic systems. A COLTRIMS reaction microscope was used to measure the squares of the vibrational wave functions of H$_2$, He$_2$, Ne$_2$, and Ar$_2$. The Schrödinger equation relates the curvature of the wave function to the potential V(R) and therefore offers a simple but elegant way to extract the shape of the potential.

# Introduction

Interaction potentials between the building blocks of matter shape the structure of bound species on a fundamental level. In the context of atomic and molecular physics potentials of interest are e.g. the van der Waals potential or, within the Born-Oppenheimer approximation, the potential energy surfaces of molecules. For a given potential the Schrödinger equation is typically considered as the condition equation for the wave function of a bound system. Mathematically, however, also the reverse is true: a given bound state wave function Ψ(R) of a two-particle system determines the full functional dependence of the interaction potential V(R) between the particles:

$$V(R) = \frac{\frac{\hbar^2}{2\mu}\frac{d^2\Psi(R)}{dR^2}}{\Psi(R)} + E \qquad (1)$$

Furthermore, the binding energy E is also contained in the wave function, as the wave function's exponential decay for $R \to \infty$ is solely determined by the reduced mass µ of the system and E [1, 2]. The wave function of a diatomic vibrational ground state is real-valued. Therefore, such wave functions and their second derivative are experimentally accessible by measuring their density distribution. In molecular physics, today, state of the art imaging techniques like Coulomb explosion imaging provide such density distributions and hence open the door to pursue this direct access to interaction potentials as we will show in this paper. We utilize the measured square of the wave function of the van der Waals-bound systems $He_2$, $Ne_2$, and $Ar_2$ as well as of the covalently bound $H_2$ in coordinate space to obtain the respective interaction potential V(R) as a function of the internuclear distance R.

The traditional way of probing the potential between two particles is by performing elastic scattering experiments. The integral and differential scattering cross-sections rely on the exact shape of the internuclear potential, which therefore can be reconstructed analyzing the measured deflection pattern. Interference patterns, originating from different trajectories leading to identical deflection angles having different phases, enable a precise determination of the attractive, the repulsive, and the well region of the potential V(R). In order to extract the shape of V(R) from such measurement, generally two approaches are possible. Firstly, one can assume a theoretical potential function between two particles and calculate the resulting deflection pattern for a given relative particle speed. This theoretical pattern is compared to the measured one, and the parameters of the function are then varied until good agreement is reached [3, 4]. Secondly, one can solve the inverse problem of scattering and infer the shape of the potential directly from the angular- and energy-dependence of the measured cross-section [5, 6]. Our alternative approach using Eq. (1) does not rely on scattering experiments at all. While the accuracy of our method is not yet competitive to accuracies reached in the established techniques, they allow, in principle, for a direct access to the shape of the potential for the full range of internuclear distances R.

## Experimental approach

We measured the square of the ground state wave functions of the diatomic systems $H_2$, $He_2$, $Ne_2$, and $Ar_2$ employing Coulomb explosion imaging using a COLTRIMS setup [7-9]. In brief, a supersonic gas jet consisting of the target molecules or rare gas dimers is intersected with ionizing light from either a synchrotron or strong femtosecond laser. Charged fragments created in the photoreaction are then guided by a homogeneous electric field towards a time and position sensitive micro channel plate detector. By measuring the positions of impact and the times of flight of each charged particle the particles' initial vector momenta can be reconstructed yielding all derived quantities as kinetic energies or emission directions. All charged photo fragments are measured in coincidence. The supersonic gas jet is created by expanding the target gas through a small aperture into a vacuum. For examining the rare gas dimers, additionally, a matter wave diffraction approach was employed as mass selector. While the expansion conditions were chosen such that considerable shares of helium, neon or argon dimers, respectively, occurred in the supersonic jet, clusters of other sizes are created, as well. In order to select the dimers from the condensed gas beam a 100 nm transmission grating has been installed in the path of the gas jet. The emerging diffraction separates particles in the gas jet according to their mass [10] towards slightly different deflection angles. The experiments were then performed by focusing the ionizing light into the part of the diffracted beam belonging to dimers.

By doubly ionizing the molecular system under investigation, two ions repelling each other by their Coulomb forces are created. In the subsequently occurring Coulomb explosion the potential energy of the two ions (initially located at an internuclear distance R) converts into kinetic energy of the ions (kinetic energy release (KER)). The KER is measured using the COLTRIMS reaction microscope and can be used to reconstruct the internuclear distance at the instant of the ionization via the so-called reflection approximation (in atomic units):

$$R = \frac{1}{KER}. \quad (2)$$

By recording a large number of Coulomb explosion events we obtain the distribution of internuclear distances R occurring in the system. Accordingly, this distribution represents a direct measurement of the square of the vibrational wave function $|\Psi(R)|^2$.

## Results and Discussion

### A. $H_2$

As a first example we present our results on the hydrogen molecule. For this experiment we used single photon double ionization by 160 eV photons provided by the Advanced Light Source (ALS) synchrotron in Berkeley (same experimental setup as in [11, 12]). At this photon energy one of the electrons absorbs the synchrotron light and the second electron is released via a shake-off or knock-off processes [13]. As Coulomb explosion imaging relies on the R-independence of the ionization process, we corrected our measured KER distributions for known small R-dependences of the respective double ionization process. The photoelectron emission from homonuclear diatomic molecules is known to be strongly influenced by interference phenomena which appear due to the two-center nature of the molecule. This interference is known to modulate the photoabsorption cross-section of the molecule [14, 15] according to

$$\sigma \sim \sigma_H(Z^*)\left(1 + \frac{\sin(kR)}{kR}\right), \quad (3)$$

where $\sigma_H$ is the atomic photoionization cross section for an effective charge $Z^*$ and k is the electron wave-vector. In single photon double ionization, as demonstrated in [12], a corresponding dependence is observable in the sum momentum of the two emitted electrons. The measured $|\Psi(R)|^2$ has been corrected correspondingly, applying a mean sum momentum $k_{mean}$ of 2.9 a.u.. Despite this inconvenience, single photon double ionization induced by soft X-ray synchrotron light is far superior to e.g. laser-based sequential double ionization for triggering a Coulomb explosion, as it is instantaneous on the time scale relevant for nuclear motion. This is especially vital for light systems with steep potential energy surfaces of the intermediate singly charged ionic state, as for example $H_2^+$.

Figure 1 shows the results of our measurement and the application of Eq. (1) to the measured data in order to obtain V(R). A binding energy $D_0$ = 4.478 eV [16] of the hydrogen molecule was assumed. After adjusting the measured KER values by 3%, the measured potential shows excellent agreement with the calculated potential energy curve [17]. Furthermore, the measured potential allows, for example, to extract the depths of the potential $D_e$ and the equilibrium distance $R_e$. Table I presents the values obtained by fitting a Morse function to the measured $H_2$ potential.

## B. Van der Waals-bound systems

The van der Waals-bound systems have been investigated using sequential tunnel ionization induced by a strong ultrashort laser pulse (Ti:Sa laser, Dragon KMLabs, 780 nm, 40 fs) in order to initiate the Coulomb explosion (see [18, 19] for a discussion of the influence of nuclear motion during the laser pulse). Also in this case we account for an R-dependence of the ionization rate. The ionization probability of tunnel ionization depends on the effective principal quantum number n* which factors in the ionization potential and, for the second ionization step, the kinetic energy released in the Coulomb explosion [20]:

$$n^* = \frac{Z}{\sqrt{2(I_P + KER)}} \quad (4)$$

The tunnel ionization rate

$$\omega_{ADK} = \sqrt{\frac{3Fn^{*3}}{\pi Z^3}} \frac{FD^2}{8\pi Z} exp\left(\frac{2Z^3}{3Fn^{*3}}\right) \qquad (5)$$

depends on the electric field of the laser F and the charge of the remaining ion Z, with $D = (4eZ^3/Fn^{*4})^{n^*}$. Since the KER depends on the distance of the atoms at the instant of ionization (see Eq. (2)), the tunnel ionization probability $\omega_{ADK}$ shows a corresponding dependency. Therefore, the distribution of R obtained in the measurement is not equal to the square of the vibrational wave function of the system, but equals the square of the wave function multiplied by tunnel ionization probabilities according to Eq. (5). The influence of this dependency can be substantial as Figs. 3 and 4 reveal. The peak intensity of the laser in the focus, which is needed when evaluating Eq. (5), was estimated by inspecting the ratio of the single ionization of monomers to the double ionization of dimers, after considering the detector efficiency of 0.6 and the fraction of dimers in the supersonic gas jet (1 % for $Ne_2$ and 5 % for $Ar_2$). This results in a laser peak intensity of $2.3 \cdot 10^{14}$ W/cm² and $8.2 \cdot 10^{13}$ W/cm² in the $Ne_2$ and $Ar_2$ experiments, respectively. The measurement of $He_2$ was conducted at a higher laser intensity, saturating the helium single ionization probability. Accordingly, the dependence of Eq. 5 on the KER can be neglected in this case. The measurement of the KER requires a careful calibration of the COLTRIMS reaction microscope. This calibration was done by resolving vibrational features occurring in the double ionization of $O_2$ and comparing them to a measurement of Lundqvist et al. [21]. This results in an KER uncertainty of less than 1%.

From the measured square of the wave function we extracted the interaction potential V(R) using Eq. (1). Obviously, this requires to analyze the second derivative of the wave function Ψ which needs to be computed numerically from the experimental data. This procedure is susceptible to statistical fluctuations, especially on the edges of the potential well where the wave function and thus our signal vanishes. To reduce this influence, the curvature was calculated including five adjacent data points instead of the minimal required three in case of the $Ne_2$ and the $Ar_2$ measurements.

Figures 2, 3, and 4 show our results for the squares of the wave functions and for the potentials of $He_2$, $Ne_2$ and $Ar_2$ respectively. For $Ar_2$ we used a binding energy of 10.5 meV [22], for $Ne_2$ a binding energy of 2.09 meV [23]. For $He_2$ the binding energy of 0.15 µeV obtained from the measured wave function itself [1] is negligible on the scale of the figure.

The same quality of agreement between the predicted potential and our measurement as in the $H_2$ case is reached for $He_2$ comparing the experimental results to theoretical calculations by Przybytek et al. [24]. For the $Ne_2$ case our measured potentials show excellent agreement with the calculations reported in [25]. We note that there are more than three orders of magnitude between the depth of the potentials of $He_2$ and $H_2$ demonstrating the versatility of our approach. Only for $Ar_2$ a deviation from the theoretically modelled curve [22] remains, even after correcting for the R-dependence of the ionization probability. We attribute this to a systematical problem of this correction, as good agreement is obtained by assuming a lower ionization intensity (i.e. altering the ionization probabilities). For internuclear distances larger than 9 a.u. the measured potential energy curve of $Ar_2$ is lower than theory predicts. In fact, it converges to the value of the binding energy E. The reason for this is that the first term on the right-hand side of Eq. (1) converges to zero for larger interatomic distances in our analysis, but not to the theoretical value of –E. This shortcoming is most likely due to the low statistics of the measured square of the wave function at large interatomic distances. The same can be seen for $Ne_2$.

Table I summarizes the derived potential depths $D_e$ and equilibrium internuclear distances $R_e$ which result from fitting Lennard-Jones functions to the rare gas dimer potentials and compares them to theory values.

## *Conclusion*

We demonstrate the extraction of the interaction potential of diatomic systems from Coulomb explosion imaging data recorded by a COLTRIMS reaction microscope. By employing an ionization process which is independent of the internuclear distance of the diatomic system, or which has a well-known dependency, the square of its wave function can be imaged. The wave function, along with the binding energy of the system, can be inserted into the Schrödinger equation, resulting in an elegant way to determine the potential V(R) over the full range of internuclear distances present in the system.

**TABLE I. Parameters as obtained from the measured V(R) distributions.**

|  | $R_{e,measured}$ [a.u.] | $R_{e,theory}$ [a.u.] | $(D_e - E)_{measured}$ [meV] | $(D_e - E)_{theory}$ [meV] |
|---|---|---|---|---|
| $H_2$ | 1.34 | 1.4 [26] | -328 | -272 [26] |
| $He_2$ | 5.7 | 5.6 [28] | -0.925 | -0.948 [27] |
| $Ne_2$ | 5.77 | 5.85 [23] | -1.3 | -1.51 [23] |
| $Ar_2$ | 7.19 | 7.12 [22] | -1.66 | -1.85 [22] |

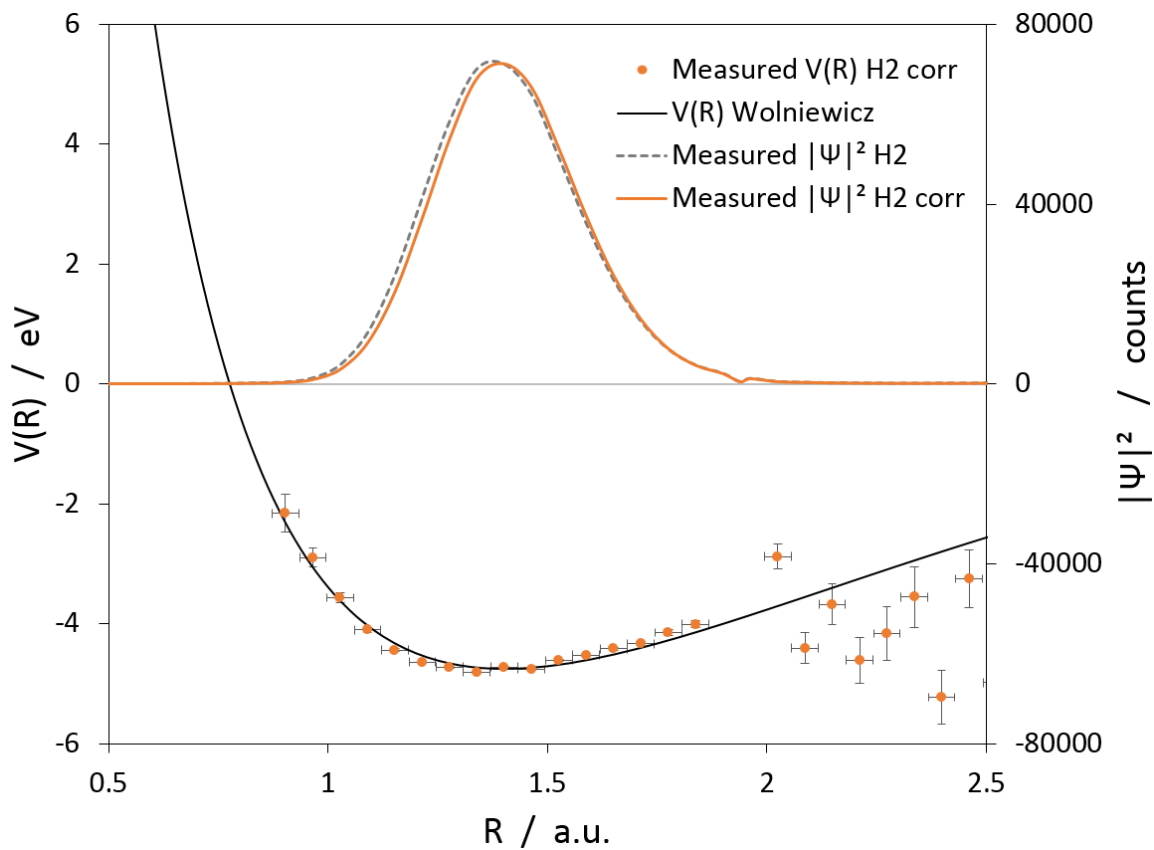

FIG. 1. Measured potential energy curve of H$_2$. Experimental data (orange dots) is shown in comparison with a calculation by Wolniewicz et al. [17] (black line). Correcting the measured distance distribution (dashed line) according to the R-dependence of the ionization probability results in the measured square of the wave function |Ψ|² (orange line).

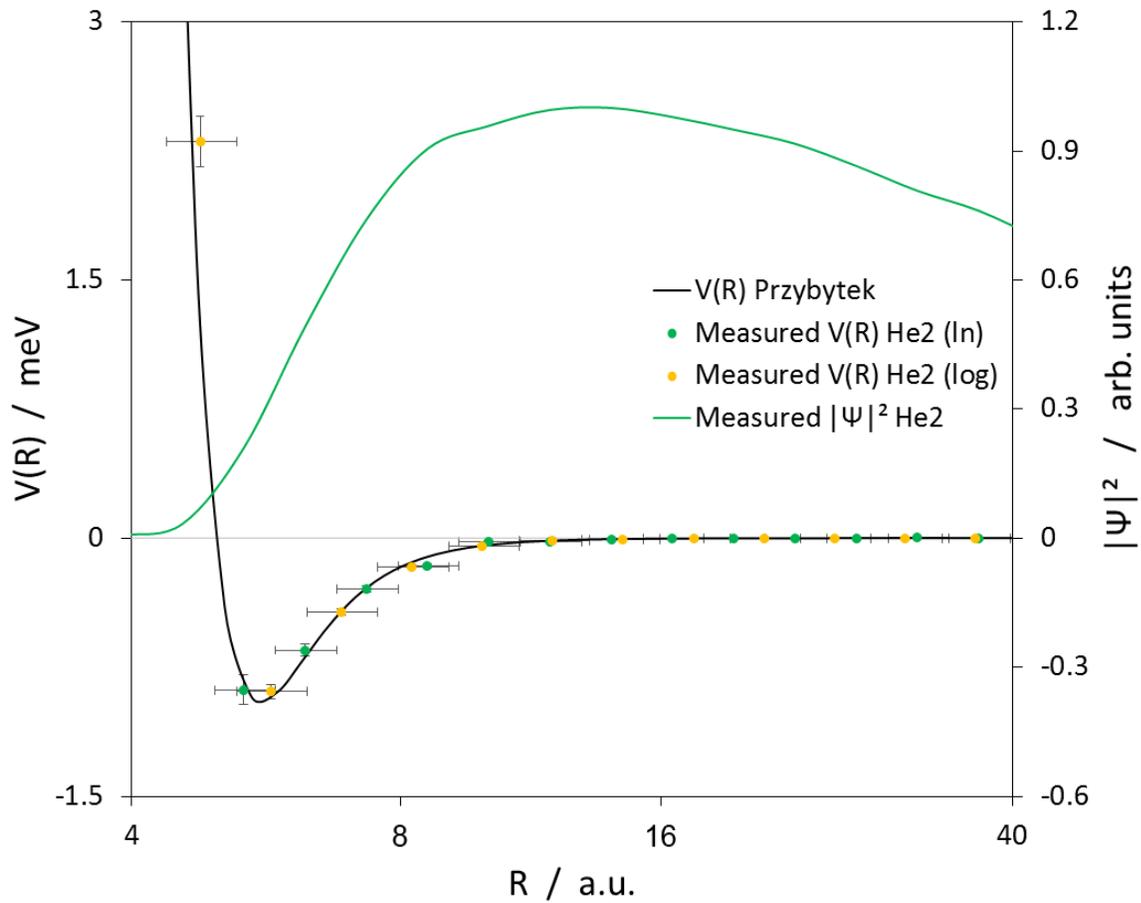

FIG. 2. Measured potential energy curve of He$_2$. Experimental data (dots) is shown in comparison with a calculation by Przybytek et al. [24] (black line). The measured square of the wave function $|\Psi|^2$ (green line) is given on a logarithmic x-scale. The green and yellow dots correspond to a binning on the basis of natural and decadic logarithm respectively.

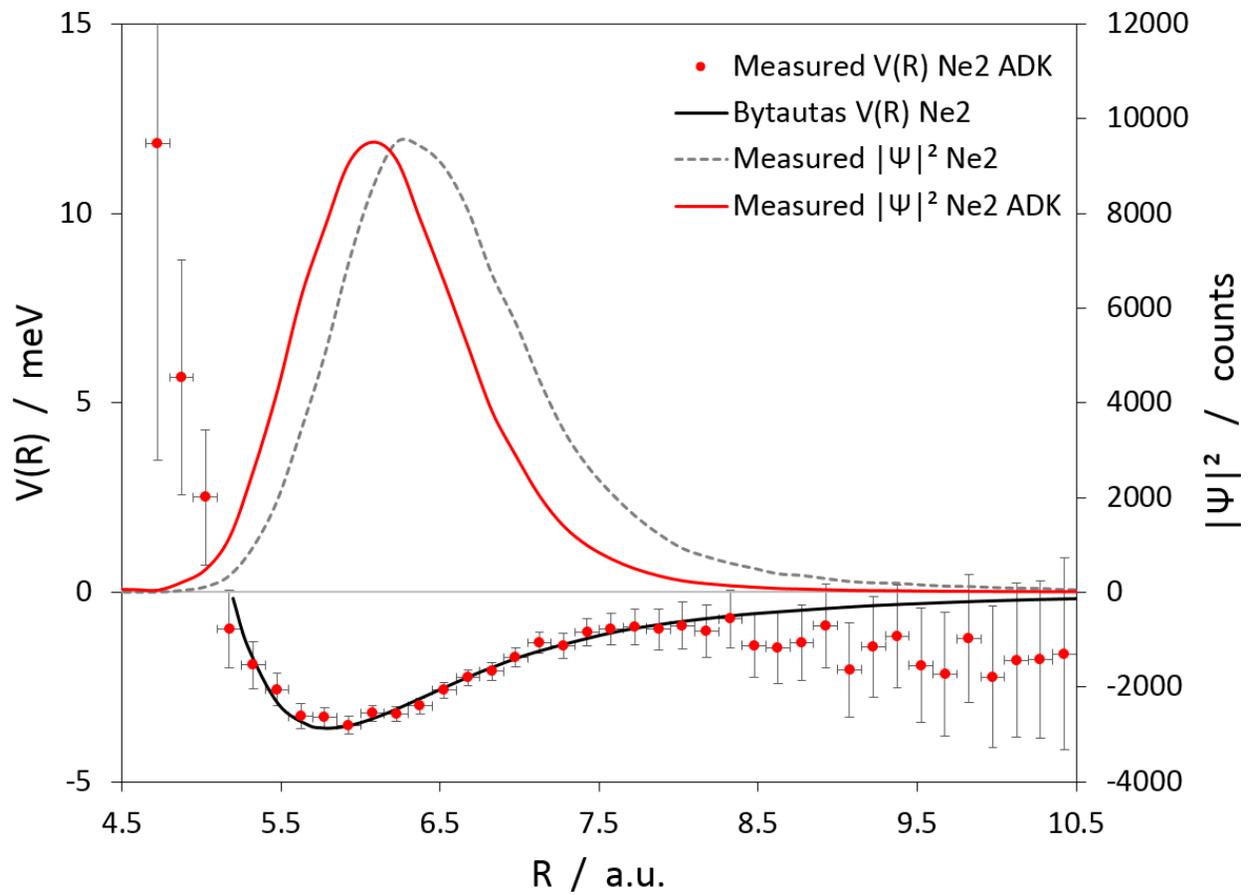

FIG. 3. Measured potential energy curve of Ne$_2$. Experimental data (red dots) is shown in comparison with a calculation (black line) by Bytautas et al. [25]. Correcting the measured distance distribution (dashed line)

according to the R-dependence of the ionization probability results in the measured square of the wave function |Ψ|² (red line).

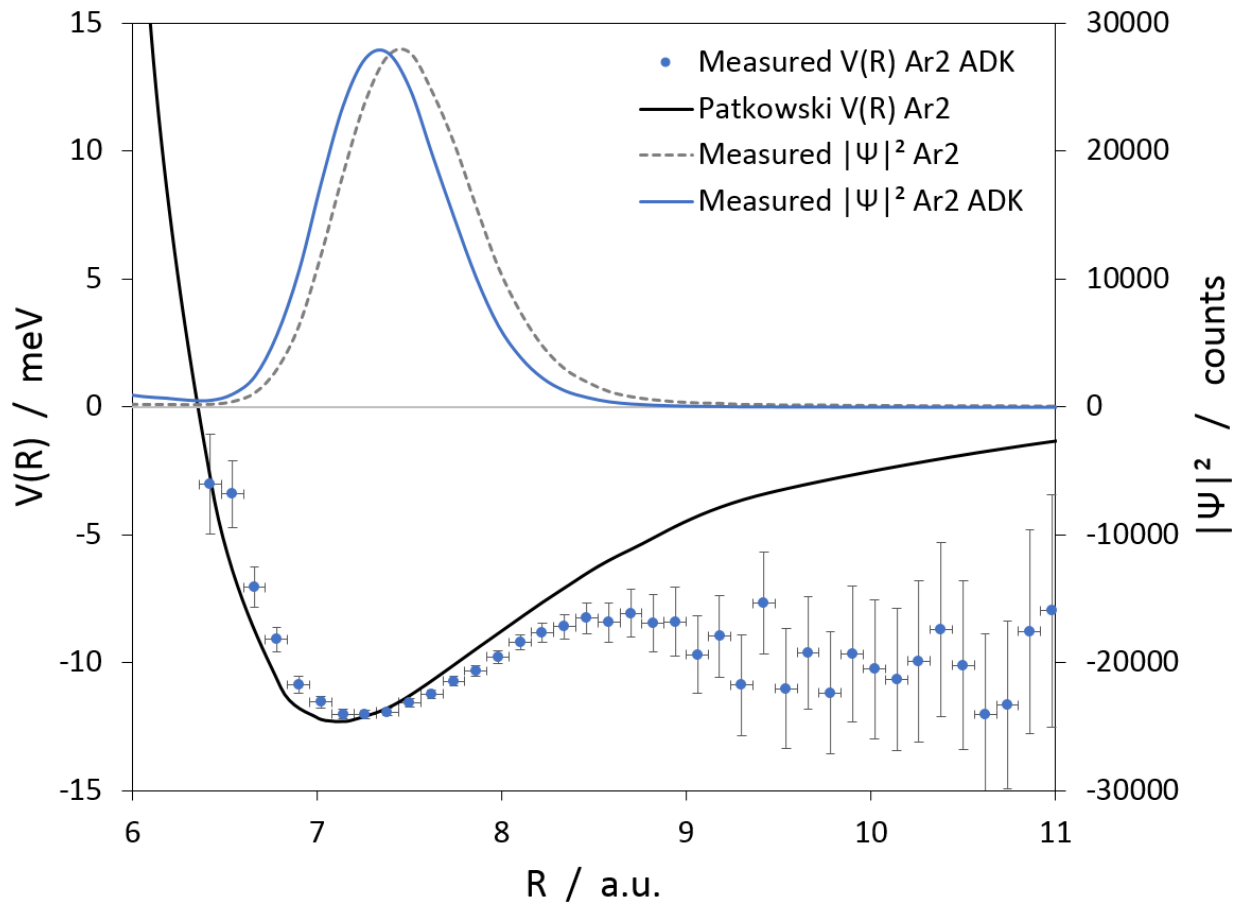

**FIG. 4. Measured potential energy curve of Ar$_2$. Experimental data (blue dots) is shown in comparison with a calculation (black line) by Patkowski et al. [22]. Correcting the measured distance distribution (dashed line) according to the R-dependence of the ionization probability results in the measured square of the wave function |Ψ|² (blue line).**

**Acknowledgements:** This work was supported by the Deutsche Forschungsgemeinschaft and the Bundesministerium für Bildung und Forschung. We thank the staff of the Advanced Light Source and of FLASH for providing the beam for part of the experiments. We thank M. Przybytek and co-workers for providing their theoretical results in numerical form. T.W. was supported by the U.S. Department of Energy Basic Energy Sciences under Contract No. DE-AC02-05CH11231.

**Author Contributions:** S.Z., M.K., J.V., and T.J. designed and constructed the experimental setup. S.Z., J.V., A.K., L.P.H.S., M.W., S.E., F.T., M.S., T.J., and R.D. performed the experiment at the femtosecond laser. A.C. contributed to the data analysis. A.K., T.J., and S.Z. developed the gas jet diffraction system. T.W. conducted the experiment at the ALS. S.Z., T.J., and R.D. analyzed the data and wrote the manuscript. All authors discussed the results and commented on the manuscript.

**Author Information:** Raw data are archived at the Goethe-University Frankfurt am Main and are available on request. Correspondence and requests for materials should be addressed to S.Z. (zeller@atom.uni-frankfurt.de) or R.D. (doerner@atom.uni-frankfurt.de).